\documentclass[11pt,a4paper]{article}

\usepackage[utf8]{inputenc}
\usepackage[T1]{fontenc}
\usepackage{fullpage}
\usepackage{amsmath,amssymb,amsthm,mathtools}
\usepackage{xcolor}
\usepackage{hyperref}

\theoremstyle{definition}

\theoremstyle{remark}

\newcommand{\dk}{d_k}
\newcommand{\dollar}{\$}
\newcommand{\suf}[1]{T_{#1}}
\newcommand{\bwt}{L}

\DeclareMathOperator{\SA}{SA}
\DeclareMathOperator{\ISA}{ISA}
\DeclareMathOperator{\LCP}{LCP}
\DeclareMathOperator{\PLCP}{PLCP}

\DeclareMathOperator{\lcp}{lcp}

\title{Substring Complexity in BWT-Runs Bounded Space}
\author{Johannes Fischer\\ TU Dortmund University, Germany\\ \texttt{johannes.fischer@cs.tu-dortmund.de}
}
\begin{document}
\maketitle

\begin{abstract}
Given a string $T[1..n]$, its substring complexity is defined as $\delta=\max_k \dk/k$, where $\dk$ is the number of its distinct substrings of length $k$ in $T$.
We show how to compute it in $O(n+r\log r)$ time, using only $O(r)$ words of space, where $r$ denotes the number of runs in the Burrows-Wheeler transform.

\end{abstract}

\section{Introduction}
\label{sec:intro}

Given a string $T[1..n]$, its substring complexity \cite{CEKNP21,RRRS13} is defined as $\delta=\max_k \dk/k$, where $\dk$ is the number of its distinct substrings in $T$ of length $k \ge 1$.
It is by now an accepted measure of the compressibility of a string \cite{Navarro21}, with many pleasant properties \cite{BCCKKP24}.
Kociumaka et al.~\cite{KNP23} show that $\delta$ (and all $\dk$'s) can be computed in $O(n)$ time from the array of longest common prefix \cite{KLAAP01}.

We show that $\delta$ can be computed in space $O(r)$ and in $O(n+r\log r)$ time, where $r$ denotes the number of runs in the Burrows-Wheeler transform.
Computing compressibility measures (such as $\delta)$ in compressed space is an important subject, as it allows us to decide if a potentially very large text is compressible or not, without using $O(n)$ space.

\subsection{Related Work}
Kawamoto and I~\cite{KI22} compute $\delta$ in $O(\rho)$ space and $O(\rho\log\log\rho)$ time from the run-length encoded version of $T$ with $\rho$ runs.
Bernardini et al.~\cite{BFGP23} compute all $\dk$ with $O(b)$ working space in $\tilde O(n^2/b)$ time for any $b$, and show a matching lower bound.
Kucherov and Nekrich~\cite{KN25} compute $\delta$ in an online manner, but use $O(n)$ space.

\subsection{Our Contribution and Techniques}
We show that $\delta$ can be computed in space $O(r)$ and in $O(n+r\log r)$ time.
The fact that $\delta  \le r$ \cite{KP18} somehow hints at the possibility of such a result.
Our algorithm relies on a relatively simple usage of \emph{irreducable} LCP-values \cite{KMP09}: they determine all other LCP-values in a regular manner, and there can be at most $r$ of them \cite{Siren10} as they are at the beginnings of BWT-runs.
The general approach is to first compute these $r$ irreducable LCP-values, which can be done in space $O(r)$ and in $O(n+r\log r)$ time \cite{SBSDNLZZ26}.
Together with their positions in the text, which can be computed in the same space, we show that any BWT-run induces at most 4 different ...

\section{Preliminaries}
Let $T=T[1..n]$ be a string over an ordered alphabet $\Sigma$ of size $\sigma$.
We assume that $T[n]=\dollar$ is a symbol that occurs nowhere else in $T$ and is smaller than every other symbol.
For $1\le i\le j\le n$, $T[i..j]$ is the substring from position $i$ to $j$, and $\suf{i}=T[i..n]$ is the $i$-th \emph{suffix}.
For strings $X$ and $Y$, $\lcp(X,Y)$ denotes the length of their longest common prefix. 
The \emph{suffix array} $\SA[1..n]$ is the permutation of $[1,n]$ with $\suf{\SA[1]}<\suf{\SA[2]}<\dots<\suf{\SA[n]}$ ~\cite{MM93}.
Its inverse is $\ISA$, $\ISA[\SA[j]]=j$.
The \emph{LCP array} is defined by $\LCP[1]=0$ and $\LCP[j]=\lcp(\suf{\SA[j]},\suf{\SA[j-1]})$ for $2\le j\le n$.
The \emph{permuted LCP array} $\PLCP$~\cite{KMP09} lists the same values in text order: $\PLCP[i]=\LCP[\ISA[i]]\qquad(1\le i\le n)$.
The \emph{Burrows-Wheeler transform} of $T$ is the string $\bwt[1..n]$ with $\bwt[j]=T[\SA[j]-1]$, where $T[0]:=T[n]=\dollar$~\cite{BW94}.
A \emph{run} in $\bwt$ is a maximal substring of equal symbols.
Let $r$ be the number of runs and $1=l_1<l_2<\dots<l_r$ their starting positions.
Let
$$
  \{u_1,u_2,\dots,u_r\}=\{\SA[l_1],\SA[l_2],\dots,\SA[l_r]\}\quad\text{with }u_1<u_2<\dots<u_r < u_{r+1}:=n+1
$$
be the sampled text positions in text order (those are the positions stored in any move-data structure \cite{BFN24}).
The intervals $I_x:=[u_x,u_{x+1})$ defined by these positions cover $T$, and make it particularly easy to compute certain PLCP-values, as for every $x\in[1,r]$ and every $i\in I_x$ we have \cite{Siren10}:
\begin{equation}
\label{eq:reducable}
\PLCP[i]=\PLCP[u_x]-(i-u_x)\ .
\end{equation}
This means that all we need to store are the PLCP-values $\PLCP[u_x]$; these are called \emph{irreducable} (the other ones are called \emph{reducable}, as they can be computed by \eqref{eq:reducable}).

For $1\le k\le n$, $\dk$ of $T$ is the number of distinct substrings $T[i..i+k-1]$ in $T$ ($1\le i\le n-k+1$), and the \emph{substring complexity} is defined as $\delta := \max_{1\le k\le n}\frac{\dk}{k}$.
Important for our algorithm is the computation of the $\dk$'s through the (P)LCP-values \cite{KNP23}:
\begin{equation}
\label{eq:dk_by_plcp}
\dk = \left|\{i\in[1,n-k+1] : \PLCP[i]<k\}\right| = (n-k+1)-\left|\{i\in[1,n] : \PLCP[i]\ge k\}\right|\ .
\end{equation}
This formula counts the contribution of every suffix to $\dk$: any PLCP-value less than $k$ means that there is a suffix not sharing the first $k$ symbols with its lexicographic predecessor, and therefore accounts +1 to $\dk$.

\section{Computation}

\begin{figure}
\includegraphics{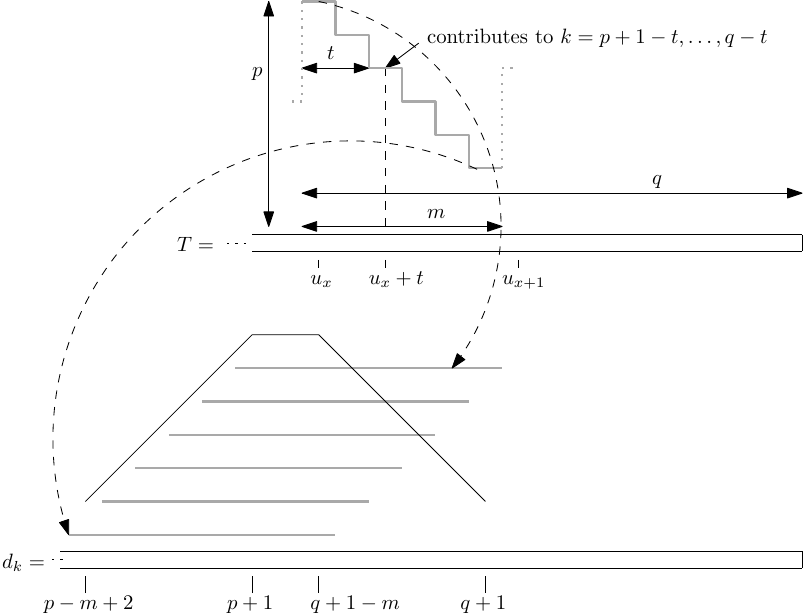}
\caption{On the top, the thicker gray line indicates the PLCP-values in the interval $I_x=[u_x, u_{x+1})$.
The suffix $\suf{u_x}$ has length $q$, and the interval $I_x$ length $m$.
On the bottom, we can see to which $k$ in $\dk$ the respective PLCP-values contribute: each to a bar of length $q-p$ (depicted in thick gray lines).
The trapezoid's height indicates the number of such bars stabbed by a position.
Note that the figure just shows the case $q-p\ge m$ such that the bottom bar ends after the top bar starts; the other case is similar.
}
\label{fig:pic}
\end{figure}

We start from \eqref{eq:dk_by_plcp}, which can also be written as
\begin{equation}
\label{eq:dk_by_plcp2}
\dk = \left|\{i :\PLCP[i]<k \le n-i+1\}\right|\ .
\end{equation}
For what follows, look also at Fig.\ \ref{fig:pic}.
Now take any interval $I_x=[u_x,u_{x+1})$, and let $p=\PLCP[u_x]$ be the PLCP-value at its beginning.
Let $q = n-u_x+1$ denote the length of $\suf{u_x}$ and look at all suffixes $\suf{u_x+t}$ starting in that interval, for $0\le t < |I_x| = u_{x+1} - u_x =: m$:
$\suf{u_x}$ contributes 1 to $\dk$ for all $k \in [p+1,q]$,
$\suf{u_x+1}$ contributes 1 to $\dk$ for all $k \in [p,q-1]$,
and so on, until
$\suf{u_{x}+m-1}$ contributes 1 to $\dk$ for all $k \in [p-m+2,q-m+1]$.
This is because \eqref{eq:reducable} gives $\PLCP[u_x+t]=\PLCP[u_x]-t$ for all $0\le t < m$, so the selection criterion in \eqref{eq:dk_by_plcp2} is equivalent to:
\begin{equation}
\label{eq:dk_by_plcp3}
\PLCP[i]<k \le n-i+1 \iff u_x+t : p-t<k \le n-(u_x+t)+1 = q-t\ .
\end{equation}

Thus, looking at $\dk$ as a function of $k$, we can say that every $u_x+t \in I_x$ contributes 1 to a size-$(q-p)$ interval in $\dk$.
Those intervals are overlapping, forming an affine function:

\begin{itemize}
  \item up to $k=p+2-m$, there is no $t$ contributing,
  \item from $k=p+1-m$ and up to $k=p+1$, there is one more contributing $t$ per increasing $k$,
  \item from $k=p+2$ and up to $k=q+1-m$, there is one more contributing $t$ per $k$, but also one less, so the number of contributing $t$'s stays the same,
  \item from $k=q+2-m$ and up to $k=q+1$, there is one less contributing $t$,
  \item and after $k=q+2$, there is again no $t$ contributing.
\end{itemize}

This means that the entire function $\dk$ is the sum of $r$ affine functions whose slopes changes only at the $4r$ extreme points where $\suf{u_x}$ and $\suf{u_{x+1}-1}$ can contribute to $\dk$ for all $x\in[1,r]$:
\begin{itemize}
  \item $p+1-m = \PLCP[u_x]+1-u_{x+1}+u_x$,
\item $p+1=\PLCP[u_x]+1$,
\item $q+1-m=n+2-u_{x+1}$, and
\item $q+1=n-u_x+2$.
\end{itemize}
In particular, the maximum $\frac{\dk}{k}$ must be attained at one of those points.

To evaluate $\dk$ at these points, we first sort them ($O(r\log r)$-time sorting suffices), and insert them into an array $B$ of length $\le 4r$, each with the value +1/-1 for the change of slopes: at $p+1-m$ and $q+1$, the slope increases by +1, as it is either the beginning of a rise, or the end of a fall; whereas at $p+1$ and $q+1-m$ a ``$-1$'' indicates a slope decrease.
We further contract $B$ by adding up all slopes at the same position, such that after that every position occurs at most once in $B$.
We can then scan $B$ from left to right, adding up the slope changes:
We start with $\dk=0$ and slope $a=0$ at $k=0$.
When coming from $y=B[i-1]$ and seeing $x=B[i]$ for some $i$, we first compute the new value $d_x = d_y + a(x-y)$, and then add the $i$'th slope value to $a$.
This allows us to keep track of the maximum ratio $\dk/k$ seen so far, hence computing $\delta$.

\bibliographystyle{alpha}
\bibliography{refs}

\end{document}